\documentclass[]{article}
\usepackage[T1, T2A]{fontenc}
\usepackage[utf8]{inputenc}
\usepackage[english, ukrainian]{babel}

\usepackage{geometry}
\usepackage{amsmath}
\usepackage{amssymb}
\usepackage{graphicx}
\usepackage{setspace}
\usepackage{titlesec}
\usepackage[]{hyperref}
\usepackage[style=numeric, sorting=none, citestyle=numeric-comp, autolang=other]{biblatex}
\usepackage{csquotes}

\DefineBibliographyStrings{ukrainian}{
  in = {},
  pages = {},
  volume = {}, 
  number = {},
  and = {\unspace,}
}

\DefineBibliographyStrings{english}{
  in = {},
  pages = {},
  volume = {}, 
  number = {},
  and = {\unspace,}
}

\DeclareFieldFormat{volume}{#1} 
\DeclareFieldFormat{number}{#1} 

\renewbibmacro*{volume+number+eid}{
  \printfield{volume}
  \iffieldundef{number}
    {}
    {\setunit{\addcomma\space}\printfield{number}}
  \setunit{\addcomma\space}
  \printfield{eid}
}

\allowdisplaybreaks

\titleformat{\section}[runin]{\normalfont\bfseries}{\thesection.}{0.2em}{}

\begin{document}

\thispagestyle{plain}
\begin{center}
    \textbf{DELAY FACTORS AND THE GENESIS OF LIMIT SETS OF THE NON-IDEAL SYSTEM ``TANK WITH LIQUID–ELECTRIC MOTOR''}   

    \vspace{0.4cm}
    \textbf{I.A. Seit-Dzhelil, A.Yu. Shvets}

    \vspace{0.4cm}    
    \textbf{ФАКТОРИ ЗАПІЗНЕННЯ Й ГЕНЕЗА ГРАНИЧНИХ МНОЖИН НЕІДЕАЛЬНОЇ СИСТЕМИ ``БАК З РІДИНОЮ–ЕЛЕКТРОДВИГУН''}
            
    \vspace{0.4cm}
    \textbf{I.А. Сеїт-Джелiль, О.Ю. Швець}
\end{center}

\centerline{\textbf{Abstact}}

Non-ideal deterministic system ``tank with liquid-electric motor'' is studied. Two delay - approximation models are considered. Impact of the delay on the emergence, evolution and disappearance of regular and chaotic limit sets (attractors) of the system are investigated. The main dynamic characteristics of the system’s steady-state regimes are computed and analyzed. Transition to chaos scenarios are studied. Realization of generalized intermittency scenario driven by delay factors is established. 

Delay factors and the genesis of limit sets of the non-ideal system ``tank with liquid-electric motor''

%Досліджується динамічна поведінка детермінованої неідеальної системи ``бак з рідиною - електродвигун'' при врахуванні факторів запізнення. Розглядаються дві моделі апроксимації запізнень. Вивчено вплив запізнень на виникнення, розвиток та зникнення граничних множин (атракторів) системи, як регулярних, так і хаотичних. Побудовані і детально проаналізовані основні динамічні характеристики усталених режимів розглянутої системи. Вивчені сценарії переходу до хаосу. Встановлена реалізація сценарію узагальненої переміжності обумовлена факторами запізнення. 

\section{Вступ.} 

Багато сучасних машин, механізмів і технічних пристроїв як конструктивні елементи мають жорсткі баки, частково заповнені рідиною. Дослідженню коливань вільної поверхні рідини у жорстких баках присвячено значну кількість робіт, теоретичних і експериментальних, серед яких можна назвати роботи: \cite{luk1, Lukovsky1980, lbk, Ibr, falt, luk2}. Особливо слід зазначити, що переважна кількість досліджень проводилася без урахування взаємодії коливальної системи, того чи іншого бака з рідиною, з джерелом збудження коливань. Такий підхід є цілком виправданим, коли потужність джерела збудження коливань  набагато перевищує потужність, що споживається коливальною системою. Якщо ж потужність джерела збудження коливань порівняна з потужністю, яку споживає коливальна система, то нехтування взаємодією коливальної системи з джерелом збудження її коливань може призвести до грубих помилок у дослідженні динамічної поведінки коливальної системи. Вперше на важливість урахування взаємодії коливальної системи с джерелом збудження коливань було вказано понад сто років тому у роботі \cite{som}. Але тільки після публікації у 1969 році монографії \cite{kon} теорія систем з обмеженим збудженням сформувалася як новий науковий напрям  математики і фізики. Системи, під час дослідження яких обов'язково враховується взаємодія коливальної підсистеми з джерелом збудження коливань, отримали назву неідеальні за Зоммерфельдом-Кононенком. Натомість системи, в яких нехтується взаємодія з джерелом збудження, називають ідеальними.

Одним з найбільших недоліків нехтування неідеальністю збудження є неправильне визначення стійкості за Ляпуновим положень рівноваги та усталених періодичних рухів коливальної системи. Так, стійкі за Ляпуновим, згідно з теоретичними розрахунками, усталені режими виявляються нестійкими при проведенні експериментів \cite{kr1}. Але значно більш серйозною проблемою дослідження різноманітних динамічних систем, в ідеальній постановці задачі, стала повна втрата інформації про реально існуючи у таких системах усталених хаотичних режимів коливань.

Особливо цікавими є випадки, коли причиною виникнення детермінованого хаосу є нелінійна взаємодія коливальної системи з джерелом збудження коливань. Для системи ``циліндричний бак, частково заповнений рідиною-електродвигун обмеженої потужності'' такі випадки розглянуті у роботах \cite{kr-sh91, kr-sh-prmech, kr-sh11, kr2, kr-sh12, sh-si, sh23}. Причому коли така система розглядається у неідеальній постановці задачі, вдається виявити велике різноманіття усталених регулярних і хаотичних граничних множин. Серед яких інваріантні тори, хаотичні і гіперхаотичні атрактори різних типів, а також хаотичні максимальні атрактори.

Різні цікаві аспекти динамічної поведінки деяких електропружних та маятникових систем  при врахуванні обмеженості джерел збудження вивчались у роботах \cite{kr-sh, be-ba, sh-kr, shonly, di-bu, sh-mac, ca-sy, li-fr, li-xu, sh-d, Donetskyi2022}. Треба відкреслити, що для цих систем були виявлені також приховані і рідкісні атрактори. Причому існування прихованих і рідкісних вдається виявити виключно завдяки розгляду взаємодії відповідної коливальної системи с джерелом збудження.

\section{Постановка задачі.} 
\begin{figure}[!ht]
  \centerline{  \includegraphics[width=0.75\textwidth]{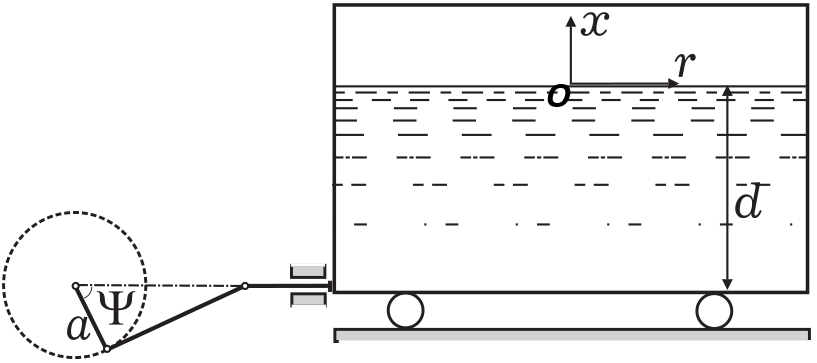}}
  \caption{Схема системи ``бак з рідиною-електродвигун''.}
  \label{fig:0}
\end{figure}

Розглянемо динамічну систему, що складається з електричного двигуна обмеженої потужності і циліндричного бака, що частково заповнений рідиною. Нехай електродвигун за допомогою кривошипно-шатунного механізму  збуджує горизонтальні коливання бака. Причому потужність, що споживається коливальним навантаженням (баком з рідиною), співмірна  з потужністю джерела збудження коливань (електродвигуна). Крім того припустимо, що швидкість обертання вала двигуна в усталеному режимі близька до власної частоти коливань вільної поверхні рідини за головними модами. Схема подібної системи наведена на Рис. \ref{fig:0}.

У випадку ідеальності джерела збудження коливання вільної поверхні рідини описуються за допомогою нелінійної крайової задачі рівняння у частинних похідних. Але резонансні коливання рідини в циліндричних за допомогою модальної системи Мійлза-Луковського \cite{luk1, Lukovsky1980, luk2, Miles1976, Miles1984b, Miles1984c} можуть бути зведені до системи звичайних диференціальних рівнянь. Явище резонансу саме по собі, як переважні коливання по одній або кількох модах, дає змогу звести дослідження континуальних систем до їх маловимірних моделей. При застосуванні модальної системи Мійлза-Луковського вдається вилучити з розгляду вплив нерезонансних мод коливань на динаміку резонансних мод. Коли ця процедура застосовується в задачах динаміки рідини в циліндрі або в сфері при резонансі, то вона виділяє, як мінімум, рівняння для двох спряжених мод, які мають однакові власні частоти і відповідають власним функціям за окружною координатою $\Psi$ вигляду $\cos{n\Psi}$ та $\sin\Psi$.
У випадку близьких власних частот, що відповідають модам з різними хвильовими параметрами, ця процедура \cite{sh08} зводить задачу до чотирьох рівнянь, причому для резонансних мод враховується їх зв’язок і взаємовплив.

У неідеальному випадку, тобто коли збудження вільних коливань рідини відбувається за допомогою електродвигуна обмеженої потужності, до системи чотирьох диференціальних рівнянь додається ще одне рівняння, яке описує взаємодію бака з рідиною та електродвигуна. Так, система диференціальних рівнянь п'ятого порядку була побудована як для випадку вимушеного, так і для випадку параметричного резонансу в роботах \cite{kr-sh-prmech, kr2}.

Зауважимо, що у попередніх дослідженнях майже не враховувався такі важливі фактори як запізнення впливів при нелінійній взаємодії бака з рідиною з джерелом збудження його коливань. Запізнення може бути присутнім у реальних коливних системах внаслідок обмеженості швидкості хвиль стиснення, розтягу, вигину, струму та електричної напруги, а також багатьох інших факторів. У деяких випадках вплив запізнення не призводить
до суттєвих змін у динамічній поведінці досліджуваних систем. В інших випадках запізнення призводить не лише до суттєвих кількісних змін характеристик усталеного руху, а й до суттєвих якісних змін типів усталених
режимів.

У роботі \cite{sd-sh} була розглянута система п'яти звичайних диференціальних рівнянь, яка описує  нелінійну взаємодію між коливаннями вільної поверхні рідини в циліндричному баку по головним резонансним модам і обертанням вала електродвигуна обмеженої потужності зі врахуванням запізнення впливу електродвигуна на коливання бака з рідиною, а також запізнення оберненого впливу коливального навантаження на функціонування джерела збудження
\begin{equation}\label{eq1}
    \begin{aligned}
        &\frac{dp_1(\tau)}{d\tau} = \alpha p_1(\tau) - [\beta(\tau - \delta) + \frac{A}{2}(p_{1}^2(\tau) + q_{1}^2(\tau) + p_{2}^2(\tau) + q_{2}^2(\tau))]q_1(\tau)+\\ &+ B(p_{1}(\tau)q_{2}(\tau) - p_{2}(\tau)q_{1}(\tau))p_2(\tau); \\
        &\frac{dq_1(\tau)}{d\tau} = \alpha q_1(\tau) + [\beta(\tau - \delta) + \frac{A}{2}(p_{1}^2(\tau) + q_{1}^2(\tau) + p_{2}^2(\tau) + q_{2}^2(\tau))]p_1(\tau)+\\ &+ B(p_{1}(\tau)q_{2}(\tau) - p_{2}(\tau)q_{1}(\tau))q_2(\tau) + 1;\\
        &\frac{d\beta(\tau)}{d\tau} = N_3 - \mu_1 q_1(\tau - \rho) + N_1\beta(\tau); \\
        &\frac{dp_2(\tau)}{d\tau} = \alpha p_2(\tau) - [\beta(\tau - \delta) + \frac{A}{2}(p_{1}^2(\tau) + q_{1}^2(\tau) + p_{2}^2(\tau) + q_{2}^2(\tau))]q_2(\tau)-\\ &- B(p_{1}(\tau)q_{2}(\tau) - p_{2}(\tau)q_{1}(\tau))p_1(\tau);\\
        &\frac{dq_2(\tau)}{d\tau} = \alpha q_2(\tau) + [\beta(\tau - \delta) + \frac{A}{2}(p_{1}^2(\tau) + q_{1}^2(\tau) + p_{2}^2(\tau) + q_{2}^2(\tau))]p_2(\tau)-\\ &- B(p_{1}(\tau)q_{2}(\tau) - p_{2}(\tau)q_{1}(\tau))q_1(\tau).
    \end{aligned}
\end{equation}
Тут фазові змінні $p_1, q_1$ та $p_2, q_2$ є амплітудами коливань вільної поверхні рідини за першою і другою основними домінантними модами відповідно; фазова змінна $\beta$ пропорційна швидкості обертання вала електродвигуна; $\tau$ — безрозмірний час; $\alpha$ — коефіцієнт в'язкого демпфування; $\mu_1$ — коефіцієнт пропорційності вібраційного моменту; $N_1$ — кут нахилу статичної характеристики електродвигуна. Параметри $A$ і $B$ є константами, що залежать від радіуса бака $R$ і висоти налитої в нього рідини $-d$; $N_3$ — мультипараметр, який залежить від радіуса бака, довжини кривошипа та власної частоти основного тону коливань вільної поверхні\cite{kr-sh-prmech, sh08, kr-sh12}. Величина $\delta$ виражає запізнення впливу електродвигуна на коливання бака з рідиною; $\rho$ -- запізнення зворотного впливу коливального навантаження на електродвигун. Причому $\rho\geq \delta\geq 0$. 

Для дослідження виникнення детермінованого хаосу у динамічних системах використовується комплекс чисельних, чисельно-аналітичних та комп'ютерних методів \cite{Kuznetsov2006, sk16}. Методика застосування цих методів для дослідження динамічної поведінки неідеальної системи ``бак з рідиною-електродвигун'' описана в \cite{kr-sh11, kr-sh12}.

\section{Спосіб апроксимації малого запізнення (Перший спосіб апроксимації).}Система \(n\) звичайних диференціальних рівнянь із запізненням аргументу є нескінченновимірною динамічною системою. Зокрема, нескінченновимірною динамічною системою буде і система рівнянь (\ref{eq1}). Тому першим кроком у дослідженнях система рівнянь (\ref{eq1}) буде наближене зведення цієї системи до скінченновимірного випадку. У випадку достатньо малих значень запізнень $\delta$ та $\rho$ може бути застосований простий та ефективний спосіб \cite{ma-si, sh-m, sh-d}. Розкладемо функції $\beta(\tau-\delta)$ та $q_1(\tau - \rho)$ у ряд Тейлора за степенями запізнень
\begin{equation}\label{eq2}
    \begin{aligned}
        \beta(\tau - \delta) &= \beta(\tau) - \delta\cdot \frac{d\beta}{d\tau} + \delta^2...\\
        q_1(\tau - \rho) &= q_{1}(\tau) - \rho \cdot \frac{dq_1}{d\tau} + \rho^2...  
    \end{aligned}
\end{equation}
Залишивши по два члени рядів у розкладах (\ref{eq2}), підставимо значення $\beta(\tau-\delta)$ та $q_1(\tau - \rho)$ у систему рівнянь (\ref{eq1}). Після перетворень отримаємо таку систему рівнянь
\begin{equation}\label{mainModel}
    \begin{aligned}
        &\frac{dp_1}{d\tau} = \alpha p_1 - [\beta - \delta(N_3 - \mu_1 q_1 + N_1\beta) + \frac{A}{2}(p_{1}^2 + q_{1}^2 + p_{2}^2 + q_{2}^2)]q_1 + B(p_{1}q_{2} - p_{2}q_{1})p_2; \\
       &\frac{dq_1}{d\tau} = \alpha q_1 + [\beta - \delta(N_3 - \mu_1 q_1 + N_1\beta) + \frac{A}{2}(p_{1}^2 + q_{1}^2 + p_{2}^2 + q_{2}^2)]p_1 + B(p_{1}q_{2} - p_{2}q_{1})q_2 + 1;\\
       &\frac{d\beta}{d\tau} = N_3 + N_1\beta  - \mu_1 q_1 + \mu_1 \rho(\alpha q_1 + [\beta + \frac{A}{2}(p_{1}^2 + q_{1}^2 + p_{2}^2 + q_{2}^2)]p_1 + B(p_{1}q_{2} - p_{2}q_{1})q_2 + 1); \\
       &\frac{dp_2}{d\tau} = \alpha p_2 - [\beta - \delta(N_3 - \mu_1 q_1 + N_1\beta) + \frac{A}{2}(p_{1}^2 + q_{1}^2 + p_{2}^2 + q_{2}^2)]q_2 - B(p_{1}q_{2} - p_{2}q_{1})p_1;\\
       &\frac{dq_2}{d\tau} = \alpha q_2 + [\beta - \delta(N_3 - \mu_1 q_1 + N_1\beta) + \frac{A}{2}(p_{1}^2 + q_{1}^2 + p_{2}^2 + q_{2}^2)]p_2 - B(p_{1}q_{2} - p_{2}q_{1})q_1.\\  
    \end{aligned}
\end{equation}

Отримана система рівнянь (\ref{mainModel}) є системою без запізнень. Розмірність фазового простору цієї системи дорівнює п'яти. Хоча параметри $\delta$ та $\rho$ не є запізненням в системі (\ref{mainModel}), ми вживатимемо цей термін і надалі, щоб підкреслити їх походження. Зазначимо, що саме такий спосіб апроксимації запізнень був застосований в роботі \cite{sd-sh}.

Для проведення чисельно-комп'ютерних експериментів оберемо значення параметрів системи (\ref{mainModel}) рівними:
\begin{equation*}
    \centering
    \begin{aligned}
        &\rho = 2\delta,~\delta \in [0; 0.03];~A = 1.12;~B = -1.531;
        &\mu_1 = 4.024;~N_3 = -1;~\alpha = -0.026;~N_1 = -1.       
    \end{aligned}
\end{equation*}

Дивергенція системи (\ref{mainModel}) $(div\,F)$ в фазовому просторі дорівнює:
\begin{equation}\label{div}  
    div\,F =  \frac{\partial(\frac{dp_1}{d\tau})}{\partial p_1} + \frac{\partial(\frac{dq_1}{d\tau})}{\partial q_1} + \frac{\partial(\frac{d\beta}{d\tau})}{\partial \beta} + \frac{\partial(\frac{dp_2}{d\tau})}{\partial p_2} + \frac{\partial(\frac{dq_2}{d\tau})}{\partial q_2} = 4\alpha + N_1 + (\delta + \rho)\mu_1p_1.
\end{equation}
Тому на відміну від систем з постійною від’ємною дивергенцією, питання про локальному зміну із плином часу фазового об'єму системи поблизу частинного розв’язку системи (\ref{mainModel}) вимагає додаткового роз'яснення. Як відомо \cite{ani}, зміну фазового об'єму в часі можна виразити як
\[V(t)=V_{0}e^{\overline{div\,F}t}\]
де $V(t)$ --- фазовий об’єм, $V(0)$ --- початковий фазовий об’єм, $\overline{div\,F}$ --- усереднена за часом дивергенція системи. Як показали проведені розрахунки для всіх, наведених далі в статті, регулярних і хаотичних атракторів усереднена по часу дивергенція системи буде від'ємною. Хоча за деяких значень часу $div\,F$ може бути додатною. Це означає, що всі атрактори системи (\ref{mainModel}) мають нульові фазові об'єми.

В якості біфуркаційного параметру виберемо запізнення $\delta$. Надійна ідентифікація типу атракторів динамічної системи може бути забезпечена тільки при сукупному дослідженні основних характеристик динамічної поведінки. Таких як спектри ляпуновських характеристичних показників, фазо-параметричні характеристики, проєкції фазових портретів і розподілів природних інваріантних мір, перерізів Пуанкаре тощо. 

Як відомо, необхідною умовою хаотичності атрактора динамічної системи є додатність максимального ляпуновського показника $\lambda_1$ \cite{Kuznetsov2006, ani, sh08}. Тому спочатку обчислимо спектр ляпуновських характеристичних показників (надалі ЛХП) для системи (\ref{mainModel}) при зміні запізнення $\delta$. Ця динамічна характеристика системи обчислювалась за допомогою узагальненого алгоритму Беннетіна-Галгані та інших \cite{be-ga76, be-ga80}. Задля побудови траєкторій розв'язків системи (\ref{mainModel}) тут і в подальшому використовувався метод  Дормана-Прінса 8(9) порядку \cite{do-pr}. Ця модифікація методу Рунге-Кутти зі змінним кроком чисельного інтегрування забезпечує надзвичайну малу похибку до $O(10^{-13})$ порядку. При відсутності запізнення у системі ($\delta=\rho=0)$ сигнатура спектра ЛХП має вигляд $\{0,-,-,-,-\}$. Тобто максимальний ЛХП $\lambda_1=0$, а всі інші ляпуновські показники --- від'ємні. Це свідчить про те, що атрактором системи є граничний цикл. Але вже при незначному збільшенні величини запізнення до $\delta \approx 0.0004$ нульовим стає і другий ляпуновський показник $\lambda_2$. Сигнатура спектру ЛХП набуває вигляду $\{0,0,-,-,-\}$. Отже атрактором системи стає інваріантний тор (квазіперіодичний атрактор).

\begin{figure}[!ht]
\centering
\includegraphics[scale=0.35]{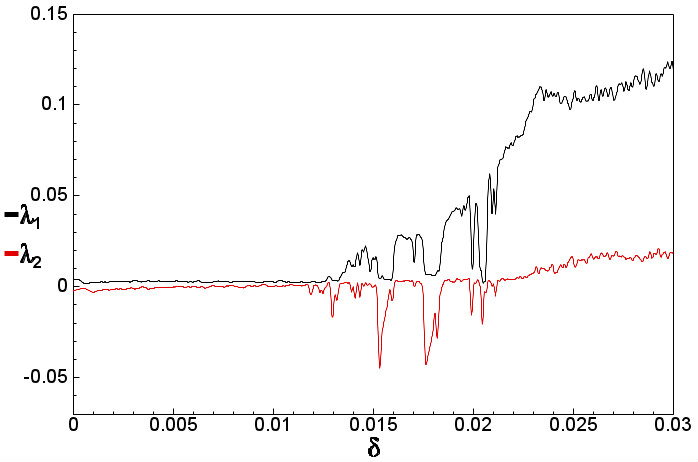}
\caption{Графік перших двох ляпуновських показників системи (\ref{mainModel}).}\label{fig:1}
\end{figure}

 На рис. \ref{fig:1} побудовано графік залежності перших двох ЛХП системи від значення запізнення $\delta$. Зауважимо, що початкові дані для проведення розрахунків вибиралися в околі початку координат фазового простору системи (\ref{mainModel}).  Крок зміни значення $\delta$ дорівнює $0.0001$. На цьому рисунку значення максимального (першого) показника нанесені чорним кольором, а другого --- червоним. Аналіз цього рисунку показує, що при $0.0004 \leq \delta \leq 0.013$ атракторами системи будуть інваріантні тори, оскільки перший (чорна лінія на графіку) і другий (червона лінія на графіку) ЛХП є нульовими (з точністю до похибки чисельного алгоритму). Зауважимо, що роздільної здатності цього рисунку на вистачає, щоб чітко ідентифікувати граничні цикли на надзвичайно малому проміжку $0\leq \delta\leq 0.0004$. Але зробити таку ідентифікацію можливо на підставі безпосереднього аналізу даних чисельних розрахунків на цьому проміжку.

 Як випливає з аналізу графіків перших двох ляпуновських показників, побудованих на рис. \ref{fig:1}, при подальшому зростанні значення $\delta$ відбувається низка біфуркацій, які призводять до зміни типів усталених режимів системи (\ref{mainModel}).

Так при $\delta > 0.013$ на відносно невеликому проміжку значень $\delta$ другий ЛХП стає від'ємним, а перший залишається нульовим. Це свідчить про виникнення граничного циклу на торі. Далі при зростанні $\delta$ перший ЛХП стає додатним, а другий --- нульовим, що свідчить про виникнення хаотичного атрактору. Зауважимо, що така послідовність біфуркацій є характерною при переході до хаосу внаслідок руйнування інваріантного тору \cite{Kuznetsov2006, sk16}. Далі, при зростанні $\delta$ відбувається низка біфуркацій ``цикл'' --- ``хаос''. І, нарешті, при $\delta > 0.22$ спостерігається біфуркація ``хаос'' --- ``гіперхаос''. Виниклий гіперхаотичний атрактор має два додатних ляпуновських показника. Сигнатура спектру ЛХП гіперхаотичного атрактора має вигляд $(+, +, 0, -, -)$. Більш детально біфуркацію ``хаос'' --- ``гiперхаос'' ми розглянемо далі. 

 Фазо-параметричну характеристику системи (надалі ФПХ) можна побудувати за допомогою метода Ено \cite{hn82, hn76}. Січною поверхнею оберемо площину  $S: \beta + 1.6=0$. На підставі системи (\ref{mainModel}) будуємо допоміжну систему рівнянь (\ref{syst5}) з новим ``часом'' $S$.  
\begin{equation}\label{syst5}
    \begin{aligned}
        &\frac{dp_1}{dS} = \frac{\alpha p_1 - [\beta - \delta(N_3 - \mu_1 q_1 + N_1\beta) + \frac{A}{2}(p_{1}^2 + q_{1}^2 + p_{2}^2 + q_{2}^2)]q_1 + B(p_{1}q_{2} - p_{2}q_{1})p_2}{N_3 + N_1\beta  - \mu_1 q_1 + \mu_1 \rho(\alpha q_1 + [\beta + \frac{A}{2}(p_{1}^2 + q_{1}^2 + p_{2}^2 + q_{2}^2)]p_1 + B(p_{1}q_{2} - p_{2}q_{1})q_2 + 1)}; \\
       &\frac{dq_1}{dS} = \frac{\alpha q_1 + [\beta - \delta(N_3 - \mu_1 q_1 + N_1\beta) + \frac{A}{2}(p_{1}^2 + q_{1}^2 + p_{2}^2 + q_{2}^2)]p_1 + B(p_{1}q_{2} - p_{2}q_{1})q_2 + 1}{N_3 + N_1\beta  - \mu_1 q_1 + \mu_1 \rho(\alpha q_1 + [\beta + \frac{A}{2}(p_{1}^2 + q_{1}^2 + p_{2}^2 + q_{2}^2)]p_1 + B(p_{1}q_{2} - p_{2}q_{1})q_2 + 1)};\\
       &\frac{d\beta}{dS} = 1; \\
       &\frac{dp_2}{dS} = \frac{\alpha p_2 - [\beta - \delta(N_3 - \mu_1 q_1 + N_1\beta) + \frac{A}{2}(p_{1}^2 + q_{1}^2 + p_{2}^2 + q_{2}^2)]q_2 - B(p_{1}q_{2} - p_{2}q_{1})p_1}{N_3 + N_1\beta  - \mu_1 q_1 + \mu_1 \rho(\alpha q_1 + [\beta + \frac{A}{2}(p_{1}^2 + q_{1}^2 + p_{2}^2 + q_{2}^2)]p_1 + B(p_{1}q_{2} - p_{2}q_{1})q_2 + 1)};\\
       &\frac{dq_2}{dS} = \frac{\alpha q_2 + [\beta - \delta(N_3 - \mu_1 q_1 + N_1\beta) + \frac{A}{2}(p_{1}^2 + q_{1}^2 + p_{2}^2 + q_{2}^2)]p_2 - B(p_{1}q_{2} - p_{2}q_{1})q_1}{N_3 + N_1\beta  - \mu_1 q_1 + \mu_1 \rho(\alpha q_1 + [\beta + \frac{A}{2}(p_{1}^2 + q_{1}^2 + p_{2}^2 + q_{2}^2)]p_1 + B(p_{1}q_{2} - p_{2}q_{1})q_2 + 1)};\\
       &\frac{d\tau}{dS} = \frac{1}{N_3 + N_1\beta  - \mu_1 q_1 + \mu_1 \rho(\alpha q_1 + [\beta + \frac{A}{2}(p_{1}^2 + q_{1}^2 + p_{2}^2 + q_{2}^2)]p_1 + B(p_{1}q_{2} - p_{2}q_{1})q_2 + 1)}.\\  
    \end{aligned}
\end{equation}

Далі на кожному  кроці чисельного інтегрування системи (\ref{mainModel}) обчислюємо величину $S_n$, де $n$ номер кроку чисельного інтегрування. Припустимо, що між $n$ і $n+1$ кроками чисельного інтегрування величина $S_n$ змінює знак. Як доведено у вище процитованих роботах Ено, проінтегрувавши тим же чисельним методом допоміжну систему (\ref{syst5}) тільки на одному кроці чисельного інтегрування рівному $-\triangle S_{n+1}$, ми отримаємо координати першої точки перерізу Пуанкаре. Продовжуючи цей процес знайдемо як завгодно багато точок перерізу Пуанкаре.

\begin{figure}[!ht]
\centering
\begin{minipage}{0.7\textwidth}
\centering
\includegraphics[width=\textwidth]{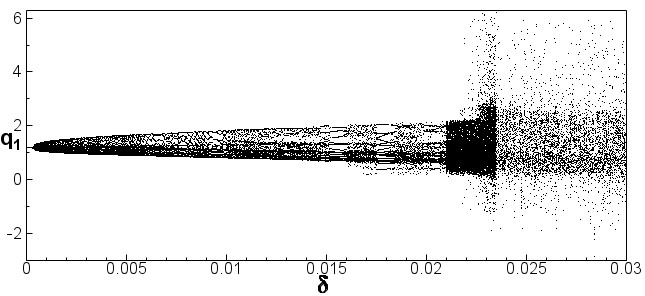}
\end{minipage}\hfill
\caption{Фазо-параметрична характеристика системи (\ref{mainModel}) за змінною  $q_1$}\label{fig:2}
\end{figure}

Фазо-параметрична характеристика  системи (\ref{mainModel}) за змінною  $ q_1$ побудована на рис. \ref{fig:2}. Зауважимо, що ця характеристика побудована для початкових значень $p_1(0)=q_1(0)=\beta(0)=p_2(0)=0, q_2(0)=1.$ Окремим лініям на цій характеристиці відповідають граничні цикли системи. Натомість густо-чорним скупченням точок відповідають або інваріантні тори, або хаотичні (гіперхаотичні) атрактори. Ідентифікація типу атрактора проводиться на підставі обчислення та аналізу сигнатури спектру ЛХП. Уважно поглянувши на рис. \ref{fig:2}, можна помітити, що при $0 \leq \delta < 0.0004$ ФПХ зображується лінією. З цього випливає, що атракторами системи є граничні цикли. Цей факт вже відзначався нами під час аналізу графіків ЛХП з рис. \ref{fig:1}, але тоді роздільної здатності цього рисунку не вистачало для чіткої візуалізації існування граничних циклів.

Отже при відсутності запізнення $(\rho=\delta=0)$ атрактором системи є граничний цикл простої однотактної структури. Але у роботі \cite{sh-si} було встановлено, що при так обраних параметрах системи (\ref{eq1}) в ній існують два граничні цикли, причому двовимірна проєкція $(q_1,q_2)$ цих циклів симетрична відносно прямої $q_2 = 0$. Звісно кожен з цих граничних циклів має власний басейн притягання. В роботі \cite{sh-si} такі симетричні граничні множини у розглянутій області значень параметрів були виявлені і для граничних циклів, і для інваріантних торів, і для хаотичних атракторів. Відмітимо, що загальна біфуркаційна поведінка, яка випливає з рис. \ref{fig:2} збігається з біфуркаційною поведінкою згаданою під час аналізу рис. \ref{fig:1}.

\begin{figure}[!ht]
\centering
\includegraphics[scale=0.55]{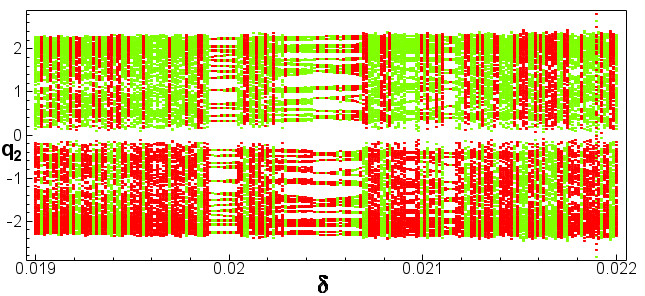}
\caption{Фазо-параметрична характеристика системи (\ref{mainModel}) за змінною $q_2.$ }\label{fig:4}
\end{figure}

Ще одним цікавим ефектом впливу різних параметрів, зокрема запізнення, на динамічну поведінку системи є так зване ``перемикання'' атракторів. На рис. \ref{fig:4} одночасно побудовані дві фазо-параметричні характеристики системи (\ref{mainModel}) за змінною $q_2$. Зеленим кольором позначені точки цієї характеристики отримані за початкових умов 
\begin{equation}
 \label{ic1} 
p_1(0)=q_1(0)=\beta(0)=p_2(0)=0, q_2(0)=-1.
\end{equation}
 При цьому запізнення $\delta$ змінювалося в інтервалі $0.019\leq\delta<0.0218$. Відповідно червоним кольором нанесені точки цієї характеристики побудовані за початкових умов
 \begin{equation}
 \label{ic2} 
p_1(0)=q_1(0)=\beta(0)=p_2(0)=0, q_2(0)=1.
\end{equation} на тому ж інтервалі зміни запізнення $\delta $. 
Вибрані початкові дані належать до різних басейнів притягання вище означених симетричних атракторів. З рис. \ref{fig:4} видно, що за фіксованого значення $\delta$ кожна з півплощин містить точки, позначені лише одним з кольорів. Проте зі зміною значення запізнення кольори точок, що належать одній з півплощин, $q_2>0$ або $q_2<0$, можуть змінюватись. Це і є ефект ``перемикання'' атракторів. Тобто при зміні запізнення траєкторія може притягнутися як до одного, так і до іншого з симетричних атракторів.  Таким чином, запізнення змінює не тільки тип атрактора (граничний цикл, інваріантний тор, хаотичний атрактор), а також може змінювати басейни притягання однотипних атракторів. Але зазначимо, що такі ``перемикання'' мають місце тільки при зміні запізнення в інтервалі $0\leq\delta<0.02185$. При подальшому збільшенні $\delta$, незалежно від початкових, траєкторії  притягуються до єдиного атрактора.

\begin{figure}[!ht]
\centering
\begin{minipage}{0.49\textwidth}
\centering
\includegraphics[width=\textwidth]{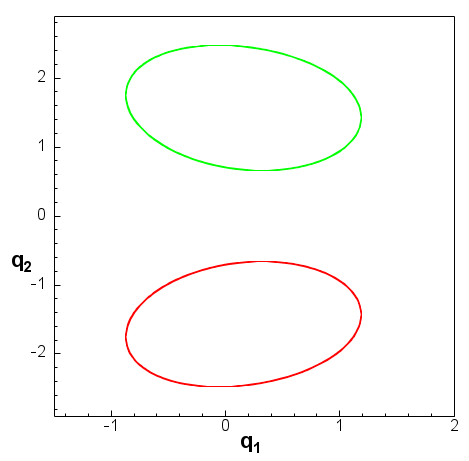}
(а) $\delta= 0$.
\end{minipage}\hfill
\begin{minipage}{0.49\textwidth}
\centering
\includegraphics[width=\textwidth]{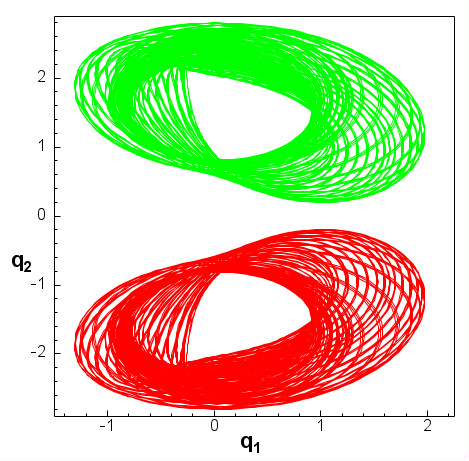}
(б) $\delta= 0.01$.
\end{minipage}
\hfill
\begin{minipage}{0.49\textwidth}
\centering
\includegraphics[width=\textwidth]{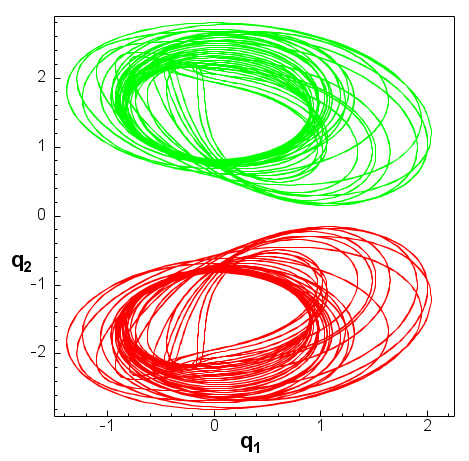}
(в) $\delta= 0.013$.
\end{minipage}\hfill
\begin{minipage}{0.49\textwidth}
\centering
\includegraphics[width=\textwidth]{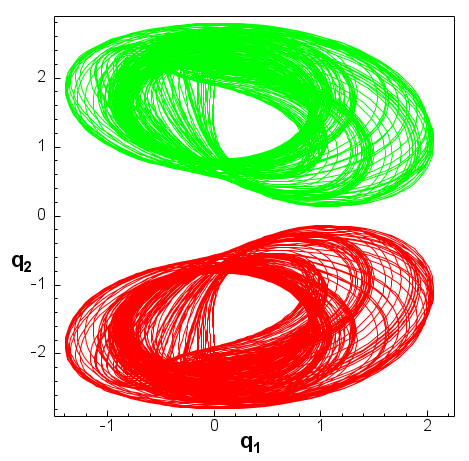}
(г) $\delta= 0.014$.
\end{minipage}
\hfill 
\caption{Проєкції фазових портретів граничних циклів (а), інваріантних торів (б), резонансних циклів на торі (в) та хаотичних атракторів (г) системи (\ref{mainModel}).}\label{fig:3}
\end{figure}

Далі побудуємо низку фазових портретів системи (\ref{mainModel}). Домовимося проєкції фазових портретів, побудованих при початкових умовах $(\ref{ic1})$ позначати зеленим кольором, а відповідні проєкції побудовані при початкових умовах $(\ref{ic2})$ позначати червоним кольором. На рис. \ref{fig:3}(a) побудовані проєкції симетричних граничних циклів, які існують у системі при відсутності запізнення. Обидва цикли мають просту однотактну структуру. У обох циклів сигнатура спектра ЛХП має вигляд $\{0,-,-,-,-\}$. Такі граничні цикли існують при $\delta\in[0,0.0004)$. При $\delta\approx 0.0004$ відбувається біфуркація Неймарка \cite{ot2002}, в результаті якої симетричні граничні цикли зникають і атракторами системи стають два симетричних інваріантних тори. Проєкції фазових портретів таких торів побудовані на рис. \ref{fig:3}(б). Сигнатури спектрів  ЛХП цих торів має вигляд $\{0,0,-,-,-\}$. Також зазначимо, що для всіх початкових даних значення ЛХП розглянутих під час досліджень атракторів збігаються з точністю до $O(10^{-4})$.

При подальшому зростанні значень запізнення у системі (\ref{mainModel}) були виявлені 
 симетричні резонансні цикли на поверхні інваріантних торів та симетричні хаотичні атрактори. Приклади проєкцій таких резонансних циклів і хаотичних атракторів наведені на рис. \ref{fig:3} (в), (г). У підрисуночних підписах вказані значення запізнень при яких побудовані ці усталені граничні множини. Зазначимо, що всі резонансні цикли на торах мають багатотактну структуру подібну приведеному на рис. \ref{fig:3} (в). При цьому сигнатура спектрів ЛХП залишається вигляду  $\{0,-,-,-,-\}$. Всі хаотичні атрактори, які існують у системи (\ref{mainModel}) при значеннях $\delta<0.0218$ мають сигнатуру $\{+,0,-,-,-\}$. 

Первинний аналіз побудованих графіків перших двох ляпуновських характеристичних показників (рис. \ref{fig:1}) і ФПХ (рис. \ref{fig:2}, рис. \ref{fig:4}) показує, що зі зростанням значення запізнень в системі (\ref{eq1}) відбуваються чисельні зміни типів атракторів. З'являються і зникають граничні цикли, інваріантні тори та хаотичні атрактори. Отже фактори запізнення істотно впливають на динамічну поведінку системи. Не зупиняючись детально на біфуркаціях Неймарка та сценаріях Фейгенбаума \cite{fei1,fei2} і Манневілля-Помо \cite{man-p,p-man} основну увагу приділимо біфуркації, яка відбувається при $\delta\approx 0.0218.$ Зауважимо, що перехід до хаосу у системі ``бак з рідиною---електродвигун'' за вищевказаними сценаріями достатньо детально вивчались у роботах \cite{kr-sh11,kr-sh12,sh-si}.

\begin{figure}[!ht]
\centering
\begin{minipage}{0.49\textwidth}
\centering
\includegraphics[width=\textwidth]{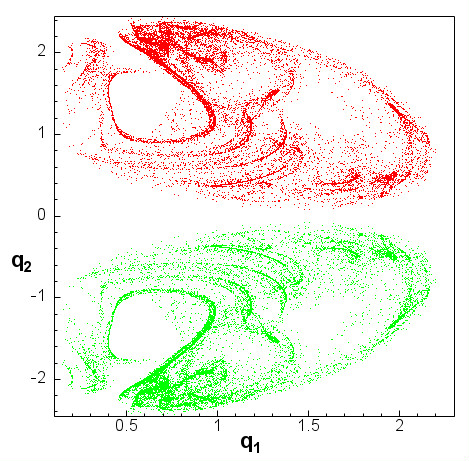}
(а) $\delta= 0.021$.
\end{minipage}\hfill
\begin{minipage}{0.49\textwidth}
\centering
\includegraphics[width=\textwidth]{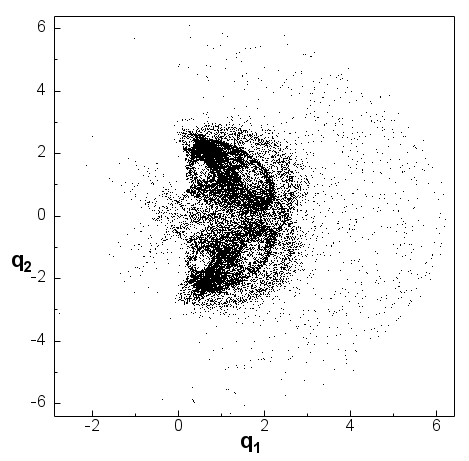}
(б) $\delta= 0.022$.
\end{minipage}
\hfill
\caption{Перерізи Пуанкаре двох симетричних хаотичних атракторів першого типу (а) та хаотичного атрактора другого типу (б) із січною площиною $ \beta +1.6=0$.}\label{fig:5}
\end{figure}

Отже у лівому півоколі точки $\delta\approx 0.0218$ існують два симетричних хаотичних атрактори системи, у кожного з яких наявний свій окремий басейн притягання. Проєкції перерізів Пуанкаре цих атракторів побудовані на рис. \ref{fig:5} (а). При збільшенні значення запізнення, а саме при $\delta > 0.0218$, два симетричних хаотичних атрактори, які існували при $0.021 < \delta < 0.0218$, зникають і у системі виникає новий атрактор. Перш за все відмітимо, що цей атрактор буде гіперхаотичним, оскільки його сигнатура спектру ЛХП має вигляд $\{+,+,0,-,-\}$. Тобто цей атрактор має не один, а два додатних ляпуновських показника. Переріз Пуанкаре виниклого гіперхаотичного атрактора побудований на рис. \ref{fig:5} (б). Перехід від двох симетричних хаотичних атракторів до гіперхаотичного відбувається за одним із варіантів сценарію узагальненої переміжності. Такий сценарій, який описує перехід від хаотичного атрактора одного типу до хаотичного атрактора іншого типу був спочатку виявлений для неідеальних динамічних систем \cite{kr-sh12, sh-si, sh21, sh-seAR}. Згодом такий сценарій був виявлений і у ідеальних динамічних системах \cite{Shvets2023, Horchakov1, Horchakov2}, і для максимальних атракторів \cite{Donetskyi, Shvets2022, Shvets205}.

На відміну від класичної переміжності за Манневіллем-Помо \cite{man-p, p-man} рух траєкторії по виниклому атрактору містить не ламінарну, а так звану грубо-ламінарну фазу. Причому цих грубо-ламінарних фаз може бути більше одної. Одними з ознак реалізації в системі сценарію узагальненої переміжності є дві наступні ознаки. По-перше, суттєве зростання об'єму області локалізації атрактора у фазовому просторі. Ця ознака чітко простежується з рис. \ref{fig:2} в околі точки $\delta\approx 0.0218$. По-друге --- збільшення величини максимального ляпуновського характеристичного показника після проходження точки біфуркації, яке можна побачити на рис. \ref{fig:1}. Виниклий гіперхаотичний атрактор являє собою ``склейку'' двох зниклих хаотичних атракторів. На рис. \ref{fig:5} (б) чітко простежуються дві густо-чорні області, які за формою дуже схожі на відповідні перерізи Пуанкаре симетричних атракторів з рис. \ref{fig:5} (а). Кожна з двох густо-чорних областей з рис. \ref{fig:5} (б) є грубо-ламінарною фазою узагальненої переміжності. Турбулентна фаза узагальненої переміжності на цьому рисунку зображається рідше нанесеними точками. Рух траєкторії по гіперхаотичному атрактору здійснюється наступним чином. Розпочавши хаотичні блукання в області першої грубо-ламінарної фази, у непередбачуваний момент часу траєкторія покидає цю область і переходить у турбулентну фазу (рідше нанесені точки на рис. \ref{fig:5} (б). Після цього, знов у непередбачуваний момент часу, траєкторія переходить в одну з грубо-ламінарних фаз. Причому можливо як повернення у першу грубо-ламінарну фазу, так і перехід у другу грубо-ламінарну фазу. Такі переходи ``одна з грубо-ламінарних фаз'' $\rightarrow$ ``турбулентна фаза'' $\rightarrow$ ``одна з грубо-ламінарних фаз'' повторюються нескінчене число разів. При цьому час знаходження траєкторії в одній з грубо-ламінарних фаз суттєво перевищує час знаходження у турбулентній фазі. Зауважимо, що реалізація такого сценарію може бути вивчена і на фазових портретах, і на розподілах природних інваріантних мір по фазовим портретам.

\section{Другий спосіб апроксимації запізнення.} Під час побудови системи (\ref{mainModel}) припускалось, що запізнення достатньо малі. Тому для апроксимації впливу запізнень були застосовані розклади в ряд Тейлора за степенями малих запізнень. Причому у цих розкладах відкидались члени другого і вищих порядків малості. Звичайно такий спосіб не підходить для дослідження динаміки системи при відносно великому значенні запізнення. У такому випадку може бути запропонована методика викладена в \cite{ma-si}. Поділимо сегмент $[-\delta;0]$ на $m$ рівних частин і введемо нові функції:
\begin{equation}
    q_{1_i}(\tau)= q_1(\tau - \frac{i\rho}{m}),~ \beta_i(\tau)= \beta(\tau - \frac{i\delta}{m}),~i\in\{0, \dots, m\}.
\end{equation}
 
Тоді систему (\ref{eq1}) можна апроксимувати за допомогою такої моделі врахування запізнення:

\begin{align*}
        &\frac{dp_1}{d\tau} = \alpha p_1 - [\beta_m + \frac{A}{2}(p_{1}^2 + q_{1}^2 + p_{2}^2 + q_{2}^2)]q_1 + B(p_{1}q_{2} - p_{2}q_{1})p_2; \\
       &\frac{dq_1}{d\tau} = \alpha q_1 + [\beta_m + \frac{A}{2}(p_{1}^2 + q_{1}^2 + p_{2}^2 + q_{2}^2)]p_1 + B(p_{1}q_{2} - p_{2}q_{1})q_2 + 1;\\
       &\frac{d\beta}{d\tau} = N_3 + N_1\beta  - \mu_1 q_{1_m}; \\
       &\frac{dp_2}{d\tau} = \alpha p_2 - [\beta_m + \frac{A}{2}(p_{1}^2 + q_{1}^2 + p_{2}^2 + q_{2}^2)]q_2 - B(p_{1}q_{2} - p_{2}q_{1})p_1;\stepcounter{equation}\tag{\theequation}\label{approximatedModel}\\
       &\frac{dq_2}{d\tau} = \alpha q_2 + [\beta_m + \frac{A}{2}(p_{1}^2 + q_{1}^2 + p_{2}^2 + q_{2}^2)]p_2 - B(p_{1}q_{2} - p_{2}q_{1})q_1;\\  
       &\frac{dq_{1_i}}{d\tau} = \frac{m}{\rho}[q_{1_{i-1}} - q_{1_i}],~i\in\{1, \dots, m\};\\
       &\frac{d\beta_i}{d\tau} = \frac{m}{\delta}[\beta_{i-1} - \beta_i],~i\in\{1, \dots, m\}.\\
\end{align*}

Система (\ref{approximatedModel}) є системою звичайних диференціальних рівнянь ($2m+5$)-ого порядку. Запізнення $\delta$ та $\rho$ є додатковими параметрами. Тут апроксимовані похідні прямують до похідних з системи (\ref{eq1}) при $m \rightarrow \infty$ \cite{ma-si}. Зауважимо, що саме такий метод  апроксимації запізнення застосовувався неідеальних динамічних систем розглянутих у роботах \cite{sh-m, sh-d}.

Аналогічно до попереднього розділу, оберемо значення параметрів рівними
\begin{equation*}
    \centering
    \begin{aligned}
        &\rho = 2\delta;~A = 1.12;~B = -1.531;
        &\mu_1 = 4.024;~N_3 = -1;~\alpha = -0.026;~N_1 = -1.       
    \end{aligned}
\end{equation*}

\begin{figure}[!ht]
\centering
\begin{minipage}{0.49\textwidth}
\centering
\includegraphics[width=\textwidth]{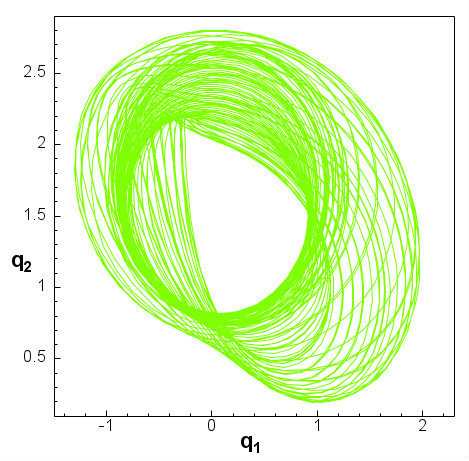}
(а) 
\end{minipage}\hfill
\begin{minipage}{0.49\textwidth}
\centering
\includegraphics[width=\textwidth]{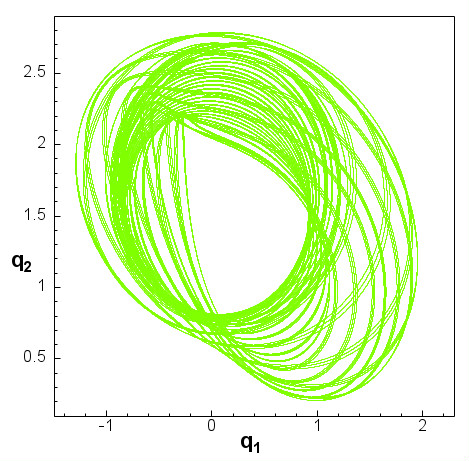}
(б) $m=50$
\end{minipage}
\hfill
\caption{Проєкції фазових портретів системи, побудовані з використанням відповідних моделей при $\delta = 0.01$.}\label{fig:6}
\end{figure}
Спочатку припустимо, що запізнення мале, наприклад, $\delta=0.01$. Перш за все проведемо порівняння результатів побудови фазових портретів за першим та другим способами апроксимації запізнення. На рис. \ref{fig:6}(a) побудована проєкція фазового портрету системи (\ref{mainModel}). Тобто розрахунки фактично проводились для вихідної системи (\ref{eq1}) за першим способом апроксимації запізнення. Відповідно на рис. \ref{fig:6}(б) побудована проєкція фазових портрету системи (\ref{approximatedModel}). Тут побудова для вихідної системи (\ref{eq1})  проводилась за другим способом апроксимації запізнення при $m=50$. Причому час необхідний для проведення комп'ютерних розрахунків за другим способом апроксимації запізнення приблизно у 50 разів перевищує час для проведення розрахунків за першим способом апроксимації запізнення. З використанням першого та другого способу апроксимації запізнення були обчислені сигнатури спектру ЛХП  для кожного з усталених режимів наведених на рис. \ref{fig:6}(a), (б). У кожному з випадків сигнатура спектра має вигляд $\{0,0,-,-,-\}$. Отже усталені режими і у одному, і у другому випадках є інваріантними торами. Зауважимо, що проєкції фазових портретів цих торів якісно збігаються. Враховуючи зауваження про неспівставні затрати часу для комп'ютерних розрахунків стає очевидним, що для малих значень запізнення другий, значно уточнений спосіб апроксимації запізнень, застосовувати недоцільно.

\begin{figure}[!ht]
\centering
\includegraphics[scale=0.55]{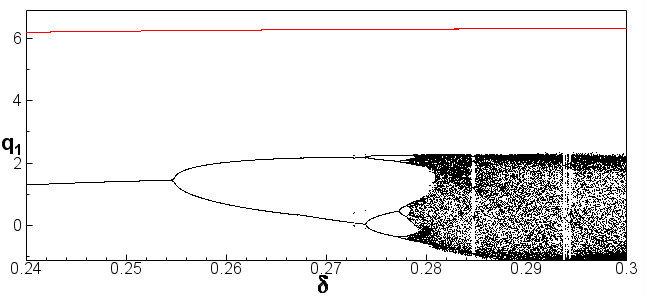}
\caption{Фазо-параметрична характеристика вихідної системи розрахована за першим способом апроксимації (червоний колір) та за другим способом апроксимації (чорний колір) із січною площиною $\beta+1.6=0$}\label{fig:7}
\end{figure}

Принципово змінюється ситуація при відносно великих значеннях запізнень. Як і у попередньому розділі в якості біфуркаційного параметру вибираємо запізнення $\delta$. Розглянемо проміжок зміни запізнення $0.24\leq\delta\leq0.3$. На рис. \ref{fig:7} побудовані ФПХ систем ``бак з рідиною --- електродвигун'' розраховані за першим способом апроксимації запізнення (нанесена червоним кольором, $q_1 > 4$) і другим способом (нанесена чорним кольором, $q_1 < 4$). Причому при другому способі апроксимації запізнень значення $m$ вибиралося з проміжку $[15,50]$. Вигляд ФПХ практично не змінювався при збільшенні $m$. Одразу стає помітною разюча несхожість цих характеристик.  ФПХ побудована за першим способом апроксимації являє собою відокремлену лінію. Така структура характеристики притаманна системам, атракторами яких є однотактні граничні цикли. Натомість ФПХ побудована за другим способом демонструє, що у досліджуваній системи відбувається складний біфуркаційний процес. Окремі гілки цієї характеристики відповідають граничним циклам, густо-чорні ділянки хаотичним атракторам. Помітні так звані вікна періодичності у хаосі. Таким чином використання простого першого способу апроксимації запізнення призводить до грубих помилок у дослідженні динаміки системи. Зокрема повністю втрачається інформація про реально існуючі хаотичні усталені режими.

\begin{figure}[!ht]
\centering
\begin{minipage}{0.49\textwidth}
\centering
\includegraphics[width=\textwidth]{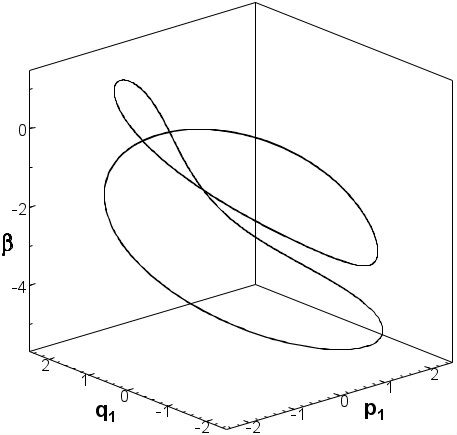}
(а) $\delta= 0.24$.
\end{minipage}\hfill
\begin{minipage}{0.49\textwidth}
\centering
\includegraphics[width=\textwidth]{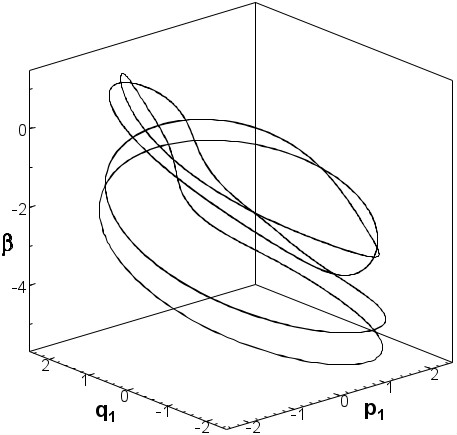}
(б) $\delta= 0.26$.
\end{minipage}
\begin{minipage}{0.49\textwidth}
\centering
\includegraphics[width=\textwidth]{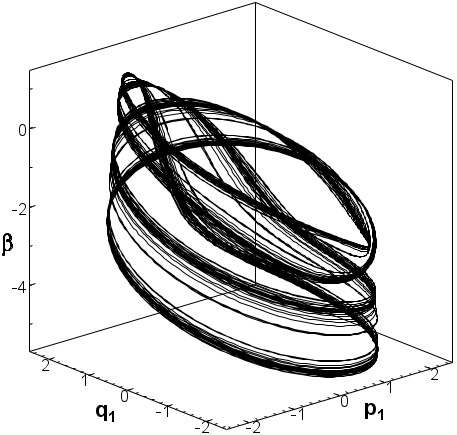}
(в) $\delta= 0.28$.
\end{minipage}
\begin{minipage}{0.49\textwidth}
\centering
\includegraphics[width=\textwidth]{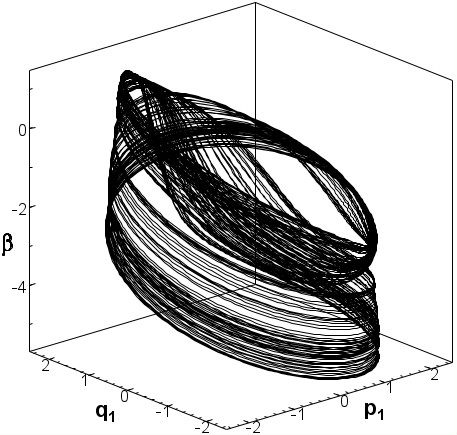}
(г) $\delta= 0.283$.
\end{minipage}
\hfill
\caption{Проєкції фазових портретів граничних циклів (а), (б) та хаотичних атракторів (в), (г).}\label{fig:8}
\end{figure}

Розглянемо деякі режими динаміки системи (\ref{eq1}) на підставі аналізу ФПХ побудованої за другим способом апроксимації запізнень. У лівій частині цієї ФПХ наявні окремі гілки, які відповідають різним  граничним циклам чітко помітні точки розгалуження гілок цього біфуркаційного дерева. Такі точки розгалуження визначають точки біфуркації граничних циклів. При $0.24\leq\delta<0.257$ єдиним атрактором системи буде однотактний граничний цикл періоду $T\approx 3.01$. Проєкція фазового портрету такого циклу побудована на рис. \ref{fig:8} (а). При зростанні запізнення, після проходження точки $\delta\approx 0.257$, цей граничний цикл втрачає стійкість і у системі народжується двотактний граничний цикл періоду $2T$. Його проєкція фазового портрету побудована на рис. \ref{fig:8} (б). Наступні біфуркації подвоєння періоду відбуваються у точках $\delta\approx 0.242$ i $\delta\approx 0.276$. У результаті цих біфуркацій у системі спочатку народжується граничний цикл періоду $4T$, а потім $8T$. Такий нескінчений каскад біфуркацій подвоєння періодів граничних циклів завершується виникненням хаотичного атрактора при $\delta\approx 0.2784$. Проєкція фазового портрету такого хаотичного атрактора, побудованого при $\delta\approx 0.28$, наведена на рис. \ref{fig:8} (в). Підтвердженням саме хаотичності цього атрактора є обчислена сигнатура спектру ЛХП, яка має вигляд $\{+,0,-,-,-\}$. Перехід від регулярного усталеного режиму до хаотичного відбувається у строгій відповідності з сценарієм Фейгенбаума \cite{fei1, fei2}.

Надалі при зростанні $\delta$ виникає новий хаотичний атрактор, проєкція фазового портрета якого побудована на \ref{fig:8} (г). Такий атрактор народжується за одну жорстку біфуркацію при проходженні точки $\delta\approx 0.2815$. Тут реалізується сценарій узагальненої переміжності \cite{sh-si, sh21}. Причому, на відміну від узагальненої переміжності знайденої в попередньому розділі, у цьому випадку ми маємо тільки одну грубо-ламінарну фазу.

\begin{figure}[!ht]
\centering
\begin{minipage}{0.49\textwidth}
\centering
\includegraphics[width=\textwidth]{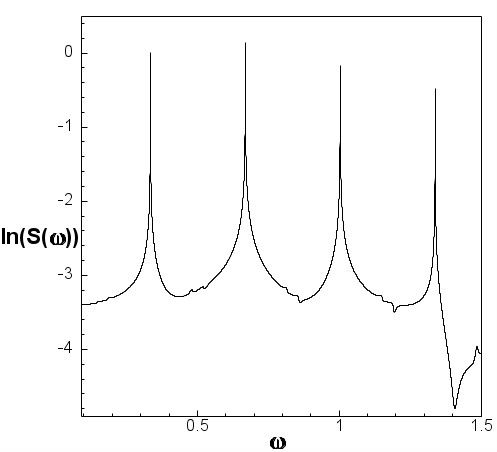}
(а) $\delta= 0.24$.
\end{minipage}\hfill
\begin{minipage}{0.49\textwidth}
\centering
\includegraphics[width=\textwidth]{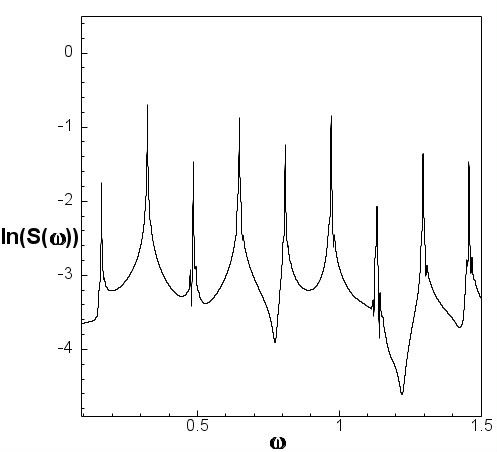}
(б) $\delta= 0.26$.
\end{minipage}
\begin{minipage}{0.49\textwidth}
\centering
\includegraphics[width=\textwidth]{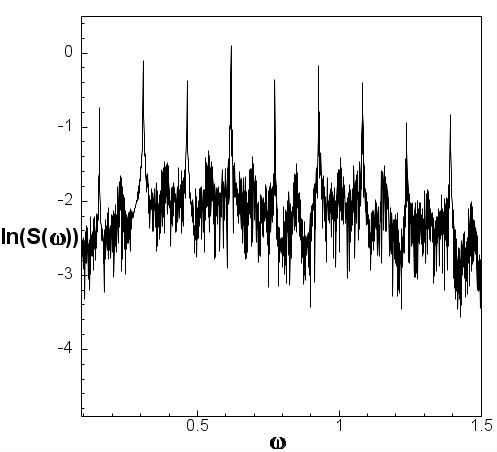}
(в) $\delta= 0.28$.
\end{minipage}
\begin{minipage}{0.49\textwidth}
\centering
\includegraphics[width=\textwidth]{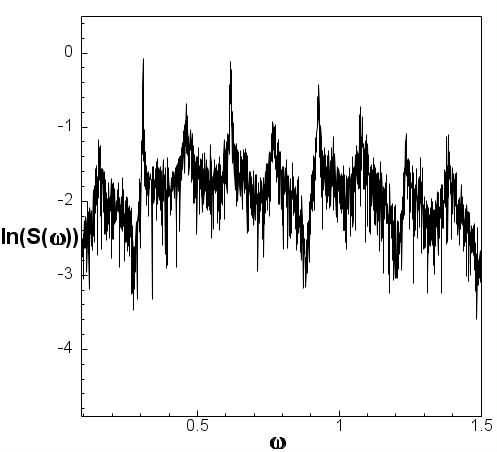}
(в) $\delta= 0.283$.
\end{minipage}
\hfill
\caption{Фур'є-спектри граничних циклів (а), (б) та хаотичних атракторів (в), (г) системи (\ref{approximatedModel}) при $m=15$.}\label{fig:9}
\end{figure}

На завершення дослідимо Фур'є-спектри побудованих атракторів неідеальної системи ``бак з рідиною --- електродвигун''. Зазначимо, що ці спектри будуються за методикою робіт \cite{SchusterJust2005, Kuznetsov2006, sh08}. Такі спектри є графіками залежності спектральної густини $S(\omega)$ від частоти $\omega$.
На рис. \ref{fig:9} (а), (б) відповідно побудовані Фур'є-спектри граничного циклу та його першої біфуркації подвоєння періоду. Обидва Фур'є-спектра граничних циклів є дискретними та рівномірними з чіткими спектральними піками. Причому після першої біфуркації подвоєння періоду можна побачити появу нових гармонік спектру  рівно посередині інтервалів між двома сусідніми гармоніками першого циклу. Поява  таких ``половинних'' гармонік спостерігається і для кожної нової біфуркації при здійсненні сценарію Фейгенбаума. На рис. \ref{fig:9} (в), (г) побудовані Фур'є-спектри двох хаотичних атракторів. Перший хаотичний атрактор виникає за сценарієм Фейгенбаума, а другий --- за сценарієм узагальненої переміжності. Обидва Фур'є-спектри неперервні, причому у другого спектру помітна тенденція до зникнення спектральних піків.

Особливо підкреслимо, що якби для досліджень при зміні значень запізнення на проміжку $0.24\leq\delta\leq 0.3$  використовувався перший, більш простий, спосіб апроксимації запізнення, то вся інформація про хаотичні усталені режими та сценарії переходу до хаосу була би втрачена.

\printbibliography

I.A. Seit-Dzhelil, National Technical University of Ukraine ``Igor Sikorsky Kyiv Polytechnic Institute'', Prospect Beresteiskyi 37, 03056, Kyiv, Ukraine \url{ilmiseitdzelil17@gmail.com}.

A.Yu. Shvets, Institute of Mathematics of the National Academy of Sciences of Ukraine, st. Tereschenkivska 3, 01024, Kyiv, Ukraine \url{oshvets@imath.kiev.ua}. 

\end{document}